\documentclass[aip,jmp,preprint,onecolumn]{revtex4-1}

\usepackage{natbib}
\usepackage{bm}
\usepackage{amsmath,amssymb,latexsym}
\usepackage{verbatim}
\usepackage{graphicx}
\usepackage{natbib}
\usepackage{subfig}
\usepackage{float}
\usepackage[usenames,dvipsnames]{color}
\usepackage{ulem} %Needed for strikeout

\restylefloat{figure}

\newcommand{\pra}{Physical Review A}
\newcommand{\prb}{Physical Review B}
\newcommand{\pre}{Physical Review E}
\newcommand{\nat}{Nature}
\newcommand{\pound}{ }

\newtheorem{theorem}{Theorem}

\begin{document}

\title{Generalized Local Induction Equation,
  Elliptic Asymptotics, and Simulating Superfluid Turbulence}
\author{Scott A. Strong and Lincoln D. Carr}
\affiliation{Department of Physics \\
  Colorado School of Mines \\ Golden, CO 80401, U.S.A.}
\date{\today}

\begin{abstract}

We prove the generalized induction equation and
the generalized local induction equation (GLIE),
which replaces the commonly used local induction
approximation (LIA) to simulate the dynamics of
vortex lines and thus superfluid turbulence. We
show that the LIA is, without in fact any
approximation at all, a general feature of the
velocity field induced by any length of a curved
vortex filament. Specifically, the LIA states that
the velocity field induced by a curved vortex
filament is asymmetric in the binormal direction.
Up to a potential term, the induced incompressible
field is given by the Biot-Savart integral, where
we recall that there is a direct analogy between
hydrodynamics and magnetostatics. Series
approximations to the Biot-Savart integrand
indicate a logarithmic divergence of the local
field in the binormal direction. While this is
qualitatively correct, LIA lacks metrics
quantifying its small parameters.  Regardless, LIA
is used in  vortex filament methods simulating the
self-induced motion of quantized vortices.  With
numerics in mind, we represent the binormal field
in terms of incomplete elliptic integrals, which
is valid for $\mathbb{R}^{3}$.  From this and
known expansions we derive the GLIE, asymptotic
for local field points. Like the LIA, generalized
induction shows a persistent binormal deviation in
the local-field but unlike the LIA,  the GLIE
provides bounds on the truncated remainder.  As an
application, we adapt formulae from vortex
filament methods to the GLIE for future use in
these methods.  Other examples we consider include
vortex rings, relevant for both superfluid $^4$He
and Bose-Einstein condensates.
 \end{abstract}

\pacs{}% insert suggested PACS numbers in braces on next line

\maketitle

\section{Introduction}

The term superfluid denotes a phase of matter whose dynamical
flows can be described, at finite non-zero temperature, by a
two-component macroscopic field with well-defined properties.
\cite{PhysRev.60.356,
1959flme.book.....L,2009JPCM...21p4220G} One component is a
purely classical field, while what remains is called the
superfluid component.  The superfluid is ideal in the sense
that it is inviscid and has infinite heat capacity provided
by its lack of classical entropy. Rotation enters the
superfluid component in quantized vortex
filaments\cite{Feynman195517,springerlink:10.1007/BF02780991}
that,  for example in $^{4}$He, transmit thermal
information acoustically and are detected by this
second-sound.  \cite{2005qvhi.book.....D}
Superfluid dynamics can be generated by the
introduction of a small heat flux.
\cite{2001LNP...571....3B} Conservation of mass
requires that the classical movement away from the
heat flux be offset by a counterflow of the
superfluid component.  If this heat flux is not
small, then the quantized vortices tangle,
indicating the onset of superfluid
turbulence.\cite{1957RSPSA.240..114V,1957RSPSA.240..128V,1957RSPSA.242..493V,1958RSPSA.243..400V,1977PhRvL..38..551S,
Schwarz1988} For large heat fluxes, the superfluid
transitions into a purely classical phase. In
classical turbulence, vorticity can concentrate
into complicated geometries. For this reason,
large-scale simulation of classical vortex
dominated flows is computationally costly.

Vortex line structures are most appropriate to superfluid
models of $^4$He where the quantized filaments have radii of
a few
angstroms.\cite{2005qvhi.book.....D,2001LNP...571...97S}
These quantized vortices provide a coherent structure for
aggressive analytical and numerical study
unavailable to classical fluids. Vortex line
structures can also be used to model atomic
Bose-Einstein condensates, where there has
recently been a revival of interest in superfluid
turbulence due in part to a series of remarkable
experiments in the Bagnato group.\cite{Henn2009}
Our study begins with
simplifications to the Biot-Savart representation
of the field induced by a vortex line
 \begin{align} \label{BS}
  \textbf{v}(\textbf{x})=
   \frac{\Gamma}{4\pi}\int_{D\subset \mathbb{R}^{3}}
   \frac{(\textbf{x}-\bm{\omega})\times d\bm{\omega}}
    {|\textbf{x}-\bm{\omega}|^{3}} =
   \frac{\Gamma}{4\pi}\int_{D\subset \mathbb{R}}
   \frac{(\textbf{x}-\bm{\xi})\times d\bm{\xi}}
    {|\textbf{x}-\bm{\xi}|^{3}}.
 \end{align}
The associated reduction of dimension aids analytic
calculation and reduces numerical cost. When such filaments
are considered initial-data to the Navier-Stokes problem,
then global well-posedness results.
\cite{Cottet1988234,Newton2001} The use of these data to
approximate self-induced vortex motion is the backbone of the
\textit{vortex filament
method}.\cite{1980JCoPh..37..289L,1985AnRFM..17..523L} The
question of numerical convergence and accuracy of such vortex
filament techniques has been addressed affirmatively in the
literature. \cite{1982MaCom..39....1B} Vortex filament
methods reduce the cost of large-scale simulations by
restricting analysis to the local field. Due to the
complexity of   classical vortical flows,
interest in these methods waned during the
1980s.\cite{1980JCoPh..37..289L,1985AnRFM..17..523L,chorin:1,Couet1981305,Ashurst1988}
However, filament methods are highly appropriate for the
constrained vortex structure associated with a superfluid.
Although (\ref{BS}) provides a straightforward starting point
for numerical computations, vortex filament methods avoid
numerical integration  altogether by replacing (\ref{BS})
with a local induction approximation.

The local induction approximation (LIA) is a result from
classical fluid dynamics, which states that a
space-curve vortex defect of an incompressible fluid field
with nontrivial curvature generates a binormal asymmetry in
the local velocity field. That is, the field local to a
length of curved vortex filament induces a
flow, which generates filament dynamics. This result, known
by Tullio Levi-Civita and his student Luigi Sante Da Rios in
the early 1900's,  \cite{1991Natur.352..561R} was
rediscovered by various post World-War II groups.
\cite{1965JFM....22..471B, 1965PhFl....8..553A,
2000ifd..book.....B, 1978SJAM...35..148C} Together, Ricca
\cite{1996FlDyR..18..245R} and Hama
\cite{1988FlDyR...3..149H} provide an excellent chronology of
LIA, a topic now common in  vortex dynamics
texts.\cite{1996QJRMS.122.1015C, 2001vif..book.....M,
Newton2001} Exploration of LIA occurs in various settings
including differential
geometry,\cite{DaRios1906,DaRios1909,DaRios1910,DaRios1911,DaRios1916a,DaRios1916b,DaRios1916c,DaRios1930,DaRios1931a,DaRios1931b,
DaRios1931c,DaRios1933a,DaRios1933b} differential
equations\cite{1965JFM....22..471B} and limits of matched
asymptotic expansions of vortex tubes.
\cite{1965PhFl....8..553A,1978SJAM...35..148C,1991JFM...222..369F}

Our derivation avoids the complications of matched asymptotic
expansions by treating vorticity concentrated to an arc.
Under this geometry Eq. (\ref{BS}) can be reduced to a
canonical elliptic representation  without the use of
power-series approximation to the  Biot-Savart
integrand.\cite{2000ifd..book.....B, 1986LPB.....4..316R}
While Taylor approximation quickly reveals binormal flow  as
a dominant feature of the induced field, it lacks error
bounds and is often restricted to a two-dimensional subspace
of $\mathbb{R}^{3}$. This paper resolves both  issues by
recasting Eq. (\ref{BS}), for a vortex-arc, into an
elliptic form valid for all field points. Using this form,
one can then use a known asymptotic formula to represent the
local field. We offer that this should be adopted, instead
of LIA, for use in vortex filament methods and Schwarz's
description of the Magnus
force.\cite{2001LNP...571...97S,Schwarz1982}
Specifically, we will prove the following results:

\begin{theorem}\label{thrm:1}
\textbf{Generalized Induction Equation}\\
Let $\bm{\omega} = \nabla \times \textbf{v}$ be localized to
an arbitrary arc with parameterization $\bm{\xi} =
(R\sin(\theta),R-R\cos(\theta),0)$, where
$R\in\mathbb{R}^{+}$ and $\theta \in D_{L}= (-L,L]$ for some
$L\in(-\pi,\pi]$. Then there exists bounded functions
$\textbf{V}_{1}$, $\alpha_{1}$,
$\alpha_{2}$, $L_{\pm}$ and $k$, of
$\varepsilon=|\textbf{x}|/R= \kappa |\textbf{x}|$, such that
the induced velocity field is given by
   \begin{align} \label{V}
    \textbf{v}(\textbf{x}) = \textbf{V}_{1}(\varepsilon)
    \left(\alpha_{1}(\varepsilon)
     \left[F(L_{+},k)-F(L_{-},k)\right] + \alpha_{2}(\varepsilon)
     \left[\frac{dF(L_{+},k)}{d\varepsilon} -
      \frac{dF(L_{-},k)}{d\varepsilon}\right]\right)
   \end{align}
where $F$ is an incomplete elliptic integral of the first
kind. Moreover, there exist constants $\beta_{1}, \beta_{2},
\beta_{3},  \beta_{4}$ such that $\textbf{V}_{1}$ can be
written as
   \begin{align}\label{V1}
    \textbf{V}_{1}(\varepsilon) =
     \varepsilon\beta_{1} \hat{\textbf{t}} -
     \varepsilon \beta_{2}\hat{\textbf{n}} +
     \left(\varepsilon\beta_{2}  +
       \varepsilon \beta_{3} +
       \beta_{4}
     \right) \hat{\textbf{b}}
    \end{align}
where $\hat{\textbf{t}},\hat{\textbf{n}},\hat{\textbf{b}},$
are the tangent, normal and binormal vectors of the local
coordinate system.
\end{theorem}

\begin{theorem}\label{thrm:2}
\textbf{Generalized Local Induction Equation}\\
Under the same hypotheses of theorem \ref{thrm:1} and for $\varepsilon
\ll 0$ the induced velocity field is dominated by the
binormal flow,
   \begin{align} \label{Ve}
     \textbf{v}_{\varepsilon}(\textbf{x})=
      4 \kappa x_{2}  \left(\alpha_{1}(\varepsilon)
     \left[F(L_{+},k)-F(L_{-},k)\right] + \alpha_{2}(\varepsilon)
     \left[\frac{dF(L_{+},k)}{d\varepsilon} -
      \frac{dF(L_{-},k)}{d\varepsilon}\right]\right)\hat{\textbf{b}}.
   \end{align}
where  $x_{2}$ is a dimensionless angular component of the
spherical decomposition of $\textbf{x}$ and $\kappa$ is the curvature of the
vortex arc. The limits $\varepsilon \to 0$ and $L \to 0$ imply that $k
\to 1$ and $\lambda \to 0$ and in this case the incomplete
elliptic integral of the first kind admits the asymptotic
relation $F \sim F_{1}$ where
  \begin{align}\label{F1}
    F_{1}(\lambda,k)=\ln\left(
      \sqrt{\frac{1+\lambda}{1-\lambda}}\right)+
     \frac{1}{\lambda}\ln\left(
     \frac{2}{1+\sqrt{(1-k^{2}\lambda^{2})/(1-\lambda^{2})}}
     \right) +\frac{1-k^{2}}{8}\ln\left(
      \frac{1+\lambda}{1-\lambda}\right).
  \end{align}
Using this, along with standard differentiation formula for
incomplete elliptic integrals of the first kind,
provides a first order  asymptotic form for the local field
given by
  \begin{align}\label{LIE}
   \textbf{v}_{\varepsilon}(\textbf{x}) &\sim
    -8\kappa x_{2} \left\{\frac{9x_{2}F_{1}(\lambda,k)}{2}-
      \frac{x_{2}E(L,k)}{(1-k^{2})k} -
      \frac{k\sin(2L)\left[\sqrt{1+k^{2}\sin^{2}(L)}+
      \sqrt{1-k^{2}\sin^{2}(L)}\right]}
      {2(1-k^{2})\sqrt{1-k^{4}\sin^{4}(L)}}\right\}
     \hat{\textbf{b}},
  \end{align}
where $E$ is an incomplete elliptic integral of the second kind.
\end{theorem}

In words, the first theorem expresses the velocity field
generated by vortex arc in terms of incomplete elliptic
integrals of the first kind. Moreover, this field can be
decomposed into three fields controlling the tangential,
circulatory and binormal flows. Of these fields, the binormal
contribution is $O(1)$ while the remaining fields are
$O(\varepsilon)$. The second theorem considers the remaining
field in the limits of $\varepsilon \to 0$ and $L \to 0$. In
this limit the incomplete elliptic integral of the first kind
admits an asymptotic form and consequently provides a
representation for the velocity field local to the vortex
arc. This asymptotic form is comparable to LIA in that the
Biot-Savart integral has been `resolved' and the binormal
flow is represented by elementary functions. This form is
only valid for filaments of infinitesimal arclength and
consequently idealized. However, using the same asymptotic
framework, one can construct expansions valid for arcs of
finite length. In fact, the remainder terms of such
expansions are known and thus the associated approximation
error can be controlled. The necessary asymptotic results are
quoted in the appendix.

The rest of this document will be organized as follows. In
Section II we define the geometry and derive the Biot-Savart
representation of the induced velocity field. In Section III
we convert this representation into an elliptic form and
prove the generalized induction result (\ref{V})-(\ref{V1}). In Section
IV we reduce this elliptic form into a sum of incomplete
elliptic integrals of the first kind. Lastly, using known
asymptotic results, we derive an expression for the local
velocity field and prove the generalized local induction
equation (GLIE) result (\ref{LIE}). We conclude with some
discussion on adapting this result to vortex filament methods
and prospective avenues of future work.

\section{Biot-Savart and Quantized Vortex Rings}

It is well known that a vortex-defect with trivial curvature
embedded into an incompressible fluid does not induce
autonomous dynamics. This is due to an angular symmetry in
the induced velocity field. This symmetry is no longer
available for curved vortex elements. Using a vortex ring, it
is possible to introduce nontrivial curvature and avoid
approximations to the Biot-Savart integral. To be precise, we
treat a vortex structure $\bm{\omega}: \mathbb{R}^{3} \to
\mathbb{R}^{3}$ such that
  \begin{align}
   \bm{\omega}(\textbf{x}) = \left\{
     \begin{array}{cc}
        1, & \textbf{x}\in \bm{\xi}\\
	0, & \textbf{x} \notin \bm{\xi}
     \end{array}\right.
  \end{align}
where $\bm{\xi}:D \to \mathbb{R}^{3}$, $D \subset
\mathbb{R}$, is parameterized by the ring
  \begin{align}
   \bm{\xi} &= R\sin(\theta) \hat{\textbf{i}} +
     \left[R-R\cos(\theta)\right]\hat{\textbf{j}},\\
   d\bm{\xi} &= \left[R\cos(\theta) \hat{\textbf{i}} +
     R\sin(\theta)\hat{\textbf{j}}\right]d\theta
   \end{align}
for $\kappa^{-1}=R\in\mathbb{R}^{+}$, $D_{L}=(-L,L]$ and
$L\in[0,\pi]$.  Thus, at the point
$\textbf{x}=(\tilde{x}_{1},\tilde{x}_{2},\tilde{x}_{3})$ we
get an element level description of the velocity field,
  \begin{align}
   v_{i}(\textbf{x}) = \int_{D_{L}}
    \frac{\epsilon_{ijk} (\tilde{x}_{j}-\xi_{j})d\xi_{k}}
    {\left[|\textbf{x}|^{2} + |\bm{\xi}|^{2}
     -2(\tilde{x}_{1}\xi_{1} +\tilde{x}_{2}\xi_{2}+
     \tilde{x}_{3}\xi_{3})
    \right]^{3/2}}
  \end{align}
where we have used the Levi-Civita symbol, $\epsilon_{ijk}$,
and employed silent-summation over repeated indices. Noting
that $|\bm{\xi}|^{2} = 2R^{2}-2R^{2}\cos(\theta)$ provides
the formulae
  \begin{align}
   v_{1}&= -|\textbf{x}|\kappa^{2}x_{3}
    \int_{D_{L}}\frac{\sin(\theta)}{D^{3/2}}d\theta,\\
   v_{2}&= |\textbf{x}|\kappa^{2}x_{3}
    \int_{D_{L}}\frac{\cos(\theta)}{D^{3/2}}d\theta,\\
   v_{3}&=\kappa^{2}|\textbf{x}|
    \int_{D_{L}}\frac{x_{1}\sin(\theta)-x_{2}\cos(\theta)}
    {D^{3/2}}d\theta + \kappa\int_{D_{L}}
    \frac{\cos(\theta)-1}{D^{3/2}}d\theta
  \end{align}
where the denominator is given by
  \begin{align}
   D&=\left[c_{1} + c_{2}\cos(\theta)+c_{3}\sin(\theta)\right]  \end{align}
and whose coefficients are
  \begin{align}
   R^{2}c_{1} &= |\textbf{x}|^{2}+2R^{2}-2\tilde{x}_{2}R,\\
   R^{2}c_{2} &= 2|\textbf{x}|{x}_{2}R-2R^{2},\\
   R^{2}c_{3} &= -2|\textbf{x}|{x}_{1}R.
  \end{align}
For future limiting work, we have chosen a
radial-representation for
$\textbf{x}=|\textbf{x}|({x}_{1},{x}_{2}, {x}_{3} )$ where
$x_{i}$ is the $i^{th}$ dimensionless angular component of
$\textbf{x}$.

\section{Conversion to Elliptic Form}

\begin{figure}[h!]
\centering
\subfloat[Vortex
Configuration]{\includegraphics[scale=1]{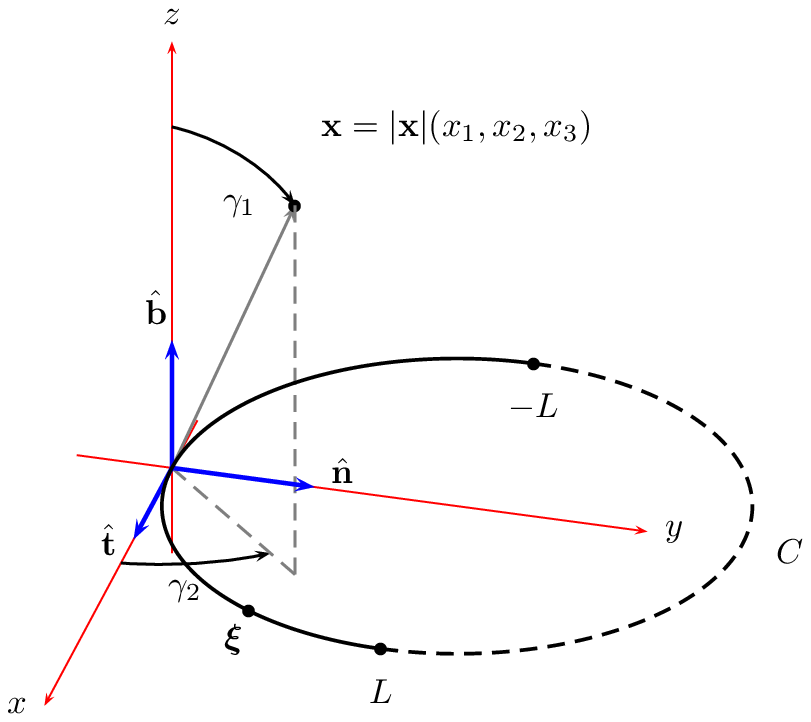}}
\subfloat[Tangent-Normal
Plane]{\includegraphics[scale=1]{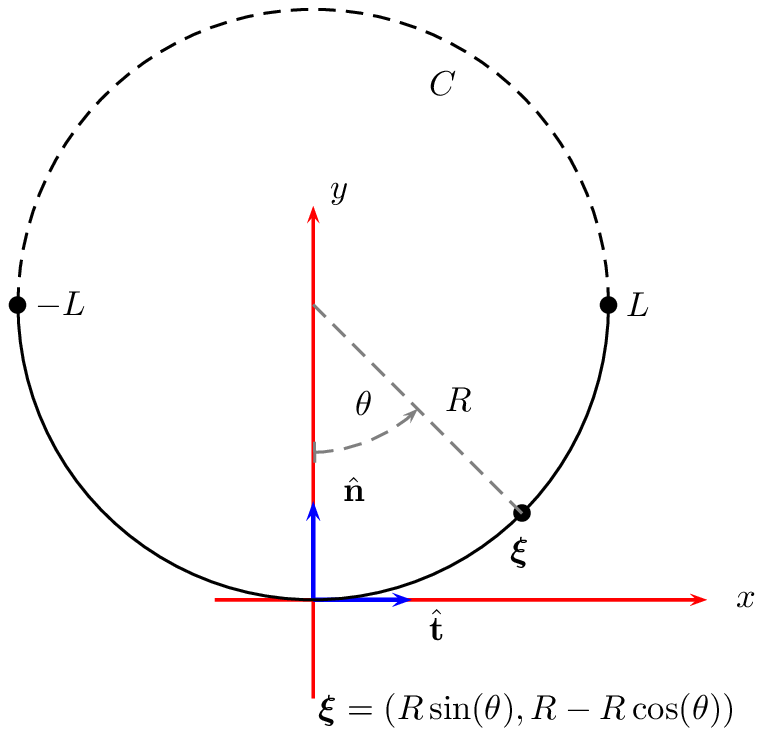}}
\caption{Global and Local Coordinate Geometry \\ In subfigure (a) the vortex
arc is depicted in $\mathbb{R}^{3}$ where the circle parameterization, $C$, is
composed of the solid line representing the vortex filament and a dashed line
representing a continuation of the parameterization. These two regions are
separated by the cut-off parameter $L$. This subfigure also shows the spherical
decomposition of the field point $\textbf{x}$ where $\gamma_{1}$ is the
azimuthal angle and $\gamma_{2}$ is the polar angle associated with the
spherical decomposition of $\textbf{x}$. Lastly, this subfigure shows the
configuration of the Serret-Frenet local basis vectors
$\hat{\textbf{t}},\hat{\textbf{n}},\hat{\textbf{b}}$,  which, for ease of use,
are oriented to correspond to the standard global basis vectors for
$\mathbb{R}^{3}$. In subfigure (b) the projection of subfigure (a) onto the
$x$--$y$ plane is given and shows  the polar decomposition of the filament
point $\bm{\xi}=(R\sin(\theta),R-R\cos(\theta),0)$.   \label{fig:1}}
\end{figure}

The previous integral representations for the velocity field
can be cast into elliptic form. To do this, we first reduce
each integral into an elliptic integral by taking derivatives
with respect to internal parameters. Doing so gives
 \begin{align}
  v_{1}&= 2\kappa x_{3}\varepsilon
   \frac{d}{dc_{3}}\int_{D_{L}}\frac{d\theta}{\sqrt{D}},\\
  v_{2}&=-2\kappa x_{3}\varepsilon
   \frac{d}{dc_{2}}\int_{D_{L}}\frac{d\theta}{\sqrt{D}},\\\
  v_{3}&= \left[2\kappa\varepsilon x_{2}
   \frac{d}{dc_{2}}-2\kappa \varepsilon x_{1}
   \frac{d}{dc_{3}}+2\kappa
   \frac{d}{dc_{1}}-2\kappa
   \frac{d}{dc_{2}}\right]\int_{D_{L}}\frac{d\theta}
   {\sqrt{D}}
 \end{align}
where the parameter $\varepsilon = |\textbf{x}|/R$ is the
ratio of radial distance to the radius of curvature.
Application of the chain-rule gives the induced velocity
field as 
  \begin{align} \label{Vint}
   \bm{v}(\textbf{x}) = \textbf{V}_{1}(\varepsilon)
    \frac{d}{d\varepsilon} \int_{D_{L}}
    \frac{d\theta}{\sqrt{D}}
  \end{align}
where the vector $\textbf{V}_{1}$ is given by
  \begin{align}
   \textbf{V}_{1}(\varepsilon) &=
    2\varepsilon \kappa x_{3}
     \frac{dc_{3}}{d\varepsilon} \hat{\textbf{i}}-
    2\varepsilon \kappa x_{2}
     \frac{dc_{2}}{d \varepsilon}\hat{\textbf{j}}+
    \left[2\kappa \varepsilon x_{2}
     \frac{dc_{2}}{d\varepsilon}-
      2\kappa \varepsilon x_{1} \frac{dc_{3}}{d\varepsilon} +
      2\kappa\frac{dc_{1}}{d\varepsilon}-
      2\kappa \frac{dc_{2}}{d\varepsilon}\right]
     \hat{\textbf{k}}
  \end{align}
implying that the velocity field is determined by the derivative of an
incomplete elliptic integral. Moreover, this proves Eq. (\ref{V1})
from theorem \ref{thrm:1} where
  \begin{align}
  \beta_{1} &= 2\kappa x_{3}\frac{dc_{3}}{d\varepsilon} ,\\
  \beta_{2} &= 2\kappa x_{2}\frac{dc_{2}}{d\varepsilon},\\
  \beta_{3} &= -2\kappa x_{1}\frac{dc_{3}}{d\varepsilon},\\
  \beta_{4} &= 2\kappa\left(\frac{dc_{1}}{d\varepsilon} -
    \frac{dc_{2}}{d\varepsilon}\right).
  \end{align}
For $\varepsilon \ll 1$ we find that the velocity field is
dominated by
  \begin{align}\label{Ve2}
   \bm{v}_{\varepsilon}(\textbf{x}) =
    -8\kappa x_{2} \frac{d}{d\varepsilon}
    \int_{D_{L}}
    \frac{d\theta}{\sqrt{D}}\hat{\textbf{k}}.
   \end{align}
Figure~\ref{fig:1} shows the vortex configuration as well as
the associated field and vortex coordinate geometry. Using
the depicted spherical decomposition of $\textbf{x}$ we find
that the previous dimensionless parameter is given by
$x_{2}=\sin(\gamma_{1})\sin(\gamma_{2})$. Moreover, we observe
that the standard basis vector $\hat{\textbf{k}}$ corresponds
to the binormal vector $\hat{\textbf{b}}$. These two facts
show that the velocity field, asymptotically close to the
vortex arc, is asymmetric in the binormal direction and that
this affect is extremized for field-points on the normal-axis, 
which agrees with standard results of induced binormal flow.

\section{Reduction of Elliptic Form to Canonical Elliptic Integrals}
Before we construct the asymptotic representation of the
velocity field, the previous integrals are converted into
canonical forms. The induced velocity field is controlled by
an integral of the form
  \begin{align}
   \int_{D_{L}} \frac{d\theta}
   {\sqrt{c_{1}+c_{2}\cos(\theta)+c_{3}\sin(\theta)}},
  \end{align}	
which can be converted to a sum of incomplete integrals of
the first kind. First, we introduce a new angle defined by
$\tan(\phi) = c_{3}/c_{2}$ and hypotenuse
$r^{2}=c_{2}^{2}+c_{3}^{2}$ to get
  \begin{align}
   \int_{D_{L}} \frac{d\theta}
    {\sqrt{c_{1}+c_{2}\cos(\theta)+
    c_{3}\sin(\theta)}}
    & =
    \int_{-L}^{L} \frac{d\theta}
     {\sqrt{c_{1}+ r \cos(\phi-\theta)}}.
  \end{align}
Now, introducing a change of variable $2\psi = \phi- \theta$
and the notation $L_{\pm} = (\phi\pm L)/2$ we apply
trigonometric formulae to get
  \begin{align}
   \int_{-L}^{L} \frac{d\theta}{\sqrt{c_{1}+
   r \cos(\phi-\theta)}}
   =
   \frac{2}{\sqrt{c_{1}+r}}\int_{L_{-}}^{L_{+}}
   \frac{d\psi}{\sqrt{1-k^{2}\sin^{2}(\psi)}}
  \end{align}
where $k^{2} = 2r/(c_{1}+r)$. Lastly,
  \begin{align}
   \int_{D_{L}} \frac{d\theta}{\sqrt{c_{1}+
    c_{2}\cos(\theta)+c_{3}\sin(\theta)}} =
    \frac{2\left[ F(L_{+},k) - F(L_{-},k)\right]}
     {\sqrt{c_{1}+r}}
  \end{align}
where $F$ is the standard incomplete elliptic integral of the
first kind,
  \begin{align} \label{elliptic}
   F(\varphi,k) = \int_{0}^{\varphi}\frac{d\psi}
    {\sqrt{1-k^{2}\sin^{2}(\psi)}} =
   \int_{0}^{\lambda} \frac{dt}
   {\sqrt{1-t^{2}}\sqrt{1-k^{2}t^{2}}}
  \end{align}
such that $\lambda = \sin(\varphi)$.

\section{Asymptotics for the Incomplete Elliptic Integral of
the First Kind}

Having reduced the Biot-Savart representation
of the velocity field to a canonical form, we can now make
use of the known asymptotic formula of Karp and
Sitnik,\cite{2007JCoAM.205..186K}  which permits the study of
(\ref{elliptic}) for all $(\lambda,k) \in [0,1]\times [0,1]$.
Specifically, they derive a series representation and
remainder term for $F$, which is asymptotic for $k \to 1$.
Their complete theorem is quoted in the appendix, but we only
require the first-order approximation
  \begin{align}
   F_{1}(\lambda, k) =  \ln\left(
   \sqrt{\frac{1+\lambda}{1-\lambda}}\right)+
   \frac{1}{\lambda}\ln\left(
    \frac{2}{1+\sqrt{(1-k^{2}\lambda^{2})/(1-\lambda^{2})}}
   \right) +\frac{1-k^{2}}{8}\ln\left(
    \frac{1+\lambda}{1-\lambda}\right).
  \end{align}
which is asymptotic to $F$ for
$\lambda \to 0$ and $k \to 1$.  This asymptotic formula is
not suited to differentiation. \cite{Erdelyi1956,
springerlink:10.1007/BF02787727,
springerlink:10.1007/BF02937348,
springerlink:10.1007/BF02795344} Thus, we must first apply
the differentiation formula, for the incomplete elliptic
integral of the first  kind, prior to its asymptotic
evaluation. Doing so gives lengthy formulae and for these we
introduce the constants
  \begin{align}
   A &= \frac{x_{2}c_{2}-x_{1}c_{3}}{r},\\
   A_{1} &= \frac{-4}{(c_{1}+r)^{3/2}}
    (\varepsilon - x_{2}+A),\\
   A_{2} &= \frac{2}{\sqrt{c_{1}+r}},\\
   A_{3} &= -2\frac{x_{2}c_{3} + x_{1}c_{2}}{r^{2}},\\
   A_{4} &= \frac{\sqrt{2r}(x_{2}-\varepsilon)}
   {(c_{1}+r)^{3/2}} + \sqrt{\frac{2}{r}}
   \left(\frac{(c_{1}+r)^{3/2}-r\sqrt{c_{1}+r}}
   {(c_{1}+r)^{2}}\right)A.
  \end{align}
Using these constants, find
  \begin{align}
   \frac{d}{d\varepsilon} \int_{D_{L}}\frac{d\theta}{\sqrt{D}}    &=
     2A_{1}[F(L_{+},k)-F(L_{-},k)] +A_{2}\left[
     \Omega(A_{3},A_{4},A_{5},k,L_{+})-
     \Omega(A_{3},A_{4},A_{5},k,L_{-})\right]
  \end{align}
where
 \begin{align}
  \Omega(A_{3},A_{4},A_{5},k,L) &=
   \frac{A_{3}}{\sqrt{1-k^{2}\sin^{2}(L)}}+
   \frac{A_{4}E(L,k)}{1(1-k^{2})k} -
   \frac{A_{4}F(L,k)}{2k} -\frac{A_{4}2k\sin(2L)}
   {4(1-k^{2})\sqrt{1-k^{2}\sin^{2}(L)}}
  \end{align}
is given by differentiation formula for incomplete elliptic integrals of
the first kind. Together with Eq. (\ref{Vint}),
proves Eq. (\ref{V})
of our first theorem where $\alpha_{1}=2A_{1}$ and $\alpha_{2}=A_{2}$.
From Eq. (\ref{Ve2}) we find that the local velocity
field is given by
 \begin{align}
 \bm{v}_{\varepsilon}(\textbf{x}) &= -8\kappa x_{2} \left(
   2A_{1}[F(L_{+},k)-F(L_{-},k)] +A_{2}\left[
   \Omega(A_{3},A_{4},A_{5},k,L_{+})-
   \Omega(A_{3},A_{4},A_{5},k,L_{-})\right]\right).
 \end{align}
At this point, the asymptotic formula (\ref{ASY}) can now be
applied to $F(\lambda,k)$ where $\lambda=\sin(L_{\pm})$. To
compare these results to standard LIA we take $\varepsilon
\to 0$ and $L \to 0$. In this case $c_{1}= -c_{2}= r\sim 2$,
$c_{3}\sim 0$ and  the constants take the 
asymptotic forms
  \begin{align}
   A &= \frac{x_{2}c_{2}-x_{1}c_{3}}{r}\sim -x_{2},\\
   A_{1} &= \frac{-4}
    {(c_{1}+r)^{3/2}}(\varepsilon - x_{2}+A)\sim x_{2},\\
   A_{2} &= \frac{2}{\sqrt{c_{1}+r}} \sim 1,\\
   A_{3} &= -2\frac{x_{2}c_{3} + x_{1}c_{2}}
    {r^{2}} \sim x_{1},\\
   A_{4} &= \frac{\sqrt{2r}(x_{2}-\varepsilon)}
    {(c_{1}+r)^{3/2}} + \sqrt{\frac{2}{r}}\left(
     \frac{(c_{1}+r)^{3/2}-r\sqrt{c_{1}+r}}
     {(c_{1}+r)^{2}}\right)A
     \sim -\frac{x_{2}}{2}.
  \end{align}
Together this gives the first-order asymptotic representation
for the velocity field,  
  \begin{align}
   \textbf{v}_{\varepsilon}(\textbf{x}) &\sim
    -8\kappa x_{2} \left\{\frac{9x_{2}F_{1}(\lambda,k)}{2}-
      \frac{x_{2}E(L,k)}{(1-k^{2})k} -
      \frac{k\sin(2L)\left[\sqrt{1+k^{2}\sin^{2}(L)}+
      \sqrt{1-k^{2}\sin^{2}(L)}\right]}
      {2(1-k^{2})\sqrt{1-k^{4}\sin^{4}(L)}}\right\}
     \hat{\textbf{b}}
  \end{align}
for the limits $\varepsilon \to 0$, $k \to1$ and $L \to 0$,
$\lambda \to 0$.  This proves Eq.
(\ref{LIE}) of our GLIE. It should be noted that the above
formula is nonzero even for the extreme case of $L=1-k\to0$. 
The physical meaning of this statement is that the
local field induced by an infinitesimal segment of a vortex
line is nonzero and asymmetric in the binormal direction.

\section{Discussion and Conclusions}

 We have derived an asymptotic representation for the local
velocity field induced by a curved vortex filament. This
derivation generalizes the previously known
statements of induced binormal flow, which play an important
role in two-component superfluid
simulation. In such simulations one must calculate the
superfluid and normal fluid flows as well as their mutual
friction interaction.  This mutual friction embodies the
scattering of rotons and phonons off of the vortex
structures.\cite{1957RSPSA.240..114V,1957RSPSA.240..128V,1957RSPSA.242..493V,1958RSPSA.243..400V,2000PhRvB..62.3409I}
It is possible to calculate this interaction  in a manner
self-consistent with Navier-Stokes and fully coupled to both
components.\cite{springerlink:10.1023/A:1004641912850} The
normal fluid is approximated through Navier-Stokes simulation
techniques while the kinematics of the superfluid make use of
LIA.  Though our focus is LIA dynamics we offer the following
references to the computational fluid dynamics literature,
which has been used for coupled two-component superfluid
simulations. \cite{1985JCoPh..59..308K,1965PhFl....8.2182H,
Wray1986} While there has been progress in these techniques,
\cite{Kennedy2000177, Ragab1997943} the recent growth of the
highly adaptable discontinuous Galerkin methods
\cite{NDG2008} and their application to nonlinear fluid flow
and acoustic problems \cite{2010JCoPh.229.6874L} is
especially provocative.

Mathematically, the kinematics of a vortex filament,
$\bm{\xi}$, are described by\cite{2005qvhi.book.....D}
  \begin{align}
   \frac{d\bm{\xi}}{dt} =
    \textbf{V}_{S}+\textbf{V}_{I}+\beta\bm{\xi}'\times
    \left(\textbf{V}_{N}-\textbf{V}_{S}-\textbf{V}_{I}\right)
    - \beta'\bm{\xi}'\times
    \left[\bm{\xi}'\times\left(\textbf{V}_{N}-
    \textbf{V}_{S}-\textbf{V}_{I}\right)\right].
  \end{align}
In a \textit{filament method}, it is typical to prescribe the
normal-fluid flow $\textbf{V}_{N}$ and neglect the mutual
friction terms involving $\beta$ and $\beta'$.
\cite{2001LNP...571...97S} This leaves only a potential flow
$\textbf{V}_{S}$ and induced flow $\textbf{V}_{I}$. Of the
remaining quantities, the computationally costly induced flow
is managed through the LIA,
  \begin{align}
   \textbf{V}_{I}(\textbf{x})\approx
    \textbf{V}_{\mathrm{local}}(\textbf{x})=
    \kappa \ln\left(\frac{2\sqrt{L_{+}L_{-}}}
    {|\textbf{x}|}\right)\hat{\textbf{b}}
  \end{align}
where $L_{\pm} = (\phi\pm L)/2$ is related to the cutoff
length $L$ and angle $\phi$. Vortex filament methods avoid
integration by application of this approximation to nodal
points of the Lagrangian computational mesh attached to the
filament centerline. Alternatively, we could simply replace
LIA with GLIE (\ref{Ve})-(\ref{F1}) and write $\textbf{V}_{I} \approx
\textbf{V}_{\varepsilon}$ and prescribe a field point
$\textbf{x}$ and arclength $s=2RL$. However, if the
higher-order circulatory and binormal terms are desired, one
could use (\ref{V})-(\ref{V1}) and employ efficient numerical routines
for the incomplete elliptic
integrals.\cite{Lemczyk1988747,ISI:A1997XL86800002,springerlink:10.1007/BF00692874}
Either of these changes will then be applied to node points
of a computational mesh modeling the filament structure.

The use of piecewise linear interpolants, while prevalent in
numerics, cause spurious effects when applied to a vortex
centerline. The interpolants themselves have zero local
curvature and their connections form cusps with undefined
local curvature. Typically, local induction is applied to
higher-order interpolations. While one can use
the generalized induction equation or GLIE
on this mesh, the natural vortex-arc
construct has been adapted to efficient meshing techniques.
\cite{Yang20021037} Consequently, computational cusps are
avoided and local curvature is always well-defined
when GLIE
is applied  to such vortex-arc meshes. Lastly, what remains
is re-meshing to allow for the experimentally witnessed
vortex nucleation.\cite{Hodby2001,2001LNP...571...36V}

Meshing is the most difficult aspect of vortex filament
implementations. Not only must the mesh adapt to the vortex
dynamics, it must be made to reconnect filament elements that
are not predicted by the Eurelian theory.
\cite{1997PhRvE..55.1617P} The most elementary reconnection
algorithms appeal to nonlinear Sch\"odinger theory and force
reconnection of filaments passing within a few core widths of
each other.\cite{Schwarz1982, 1993PhRvL..71.1375K} The
current theory of the reconnection process is not
satisfactory and efforts to avoid ad hoc simulated
reconnection
continue.\cite{2000EJMF...19..361L,2001LNP...571..177L,2001PhRvB..64u4516L,2001EL.....54..774K,
2004Nonli..17.2091G, 2010PhyD..239.1367P}

Superfluid turbulence dominated by quantized vortex flows is
an active area of  analytic, numerical and experimental
research.\cite{2010PhRvA..82c3616S,Henn2009a, Henn2009,
Henn2010, Abo-Shaeer2001, Madison2000, Kobayashi2008,
Kasamatsu2003, Castin1996, 1956RSPSA.238..204H,
1956RSPSA.238..215H, 1960AdPhy...9...89H} Though local
induction techniques will play a part in continued numerical
investigations, understanding geometric and topological
quantification of a tangled state is as important and still a
work in progress. \cite{Ricca2001, Barenghi2001, Poole2003,
Jou2010} Lastly, the vortex line approximation, while useful
and appropriate, must eventually be discarded in favor of
nontrivial core-structure. It is likely that the methods
developed within this paper can be adapted to current
arguments used to study fields induced by vortex tubes.
\cite{1965PhFl....8..553A, 1965PhFl....8..553A,
1991JFM...222..369F}

That being said, this work makes it clear that binormal
flow proportional to curvature is a general feature of vortex
filament dynamics. This means that the well-celebrated
transformation of Hasimoto\cite{Hasimoto1972}, which connects
the filament's curvature and torsion variables to a
wavefunction controlled by nonlinear Schr\"odinger evolution,
is fundamental to vortex filament dynamics. Consequently,
even geometrically complicated filament dynamics are rooted
in integrable systems theory. This connection underpins
efforts to predict allowed filament geometrics from the
associated integrable systems.
\cite{Calini97recentdevelopments,
Grinevich_closedcurves,springerlink:10.1007/s00332-004-0679-9,
1996RSPSA.452.1531U, 1992PhFl....4..938R}

\begin{acknowledgments} The authors thank Paul
Martin for useful discussions, and acknowledge
support of the National Science Foundation under
Grant PHY-0547845 as part of the NSF CAREER
program.
\end{acknowledgments}

\appendix*

\section{Asymptotic Representation for Incomplete Elliptic
Integrals of the First Kind}

The following theorem is one of the two major results proven
in Karp and Sitnik. \cite{2007JCoAM.205..186K} The second
result gives  a simpler expression but is not valid on the
leftmost edge of the unit square and therefore not used in
our calculations.

\begin{theorem}
For all $(\lambda,k)\in[0,1]\times[0,1]$ and an integer
$N\geq 1$, the previous elliptic integral admits the
representation
 \begin{align} \label{ASY}
    F(\lambda,k)&=\frac{1}{2}\ln\left(
     \frac{1+\lambda}{1-\lambda}\right)
    \sum_{j=0}^{N} \frac{(1/2)_{j}(1/2)_{j}}
     {(j!)^{2}}(1-k^{2})^{j}+ \frac{1}{2\lambda}
    \sum_{n=0}^{N-1} \left(\frac{1-\lambda^{2}}
     {-\lambda^{2}}\right)^{n} \,
    s_{n}\left(\frac{(1-k^{2})\lambda^{2}}
    {1-\lambda^{2}}\right)  + R_{N}(\lambda,k),
  \end{align}
where $s_{n}(\cdot)$ is given by the recurrence formulae
  \begin{align}
   s_{n+3}&=\frac{a_{n}s_{n+2}(x), +b_{n}s_{n+1}(x)+
    c_{n}s_{n}(x)+h_{n}}{4(n+3)^{2}},\\
   a_{n}(x) &= 8n^{2}+36n+42-x(2n+5)^{2},\\
   b_{n}(x) &= 2x(4n^{2}+14n+13)-(2n+3)^{2},\\
   c_{n}(x) &= -4x(n+1)^{2},\\
   h_{n}(x) &= \frac{x(2n+5)(2n+4)^{2}+(n+3)(8n^{2}+24n+17)}
   {8(n+3)[(n+2)!]^{2}} [(3/2)_{n}]^{2}(-x)^{n+2},\\
   s_{0}(x) &= -2 \ln \left(\frac{1+\sqrt{1+x}}{2}\right),\\
   s_{1}(x) &= \left(\frac{x}{2}-1\right) \ln \left(
     \frac{1+\sqrt{1+x}}{2}\right)-
    \frac{1}{2}\sqrt{1+x}+\frac{1}{2}+\frac{x}{2},\\
    s_{2}(x) &= \left(-\frac{9}{32}x^{2}+\frac{x}{4}-
    \frac{3}{4}\right)
    \ln \left(\frac{1+\sqrt{1+x}}{2}\right)+
     \left(\frac{9}{32}x- \frac{7}{16}\right)
    \sqrt{1+x} +\frac{7}{16} + \frac{1}{8}x-
    \frac{21}{64}x^{2}
  \end{align}
and the remainder term is negative  and satisfies,
  \begin{align}
    \frac{[(1/2)_{N+1}]^{2}(1-k^{2})^{N}}
     {2[(N+1)!]^{2}}f_{N+1}(\lambda,k)\leq
    -R_{N}(\lambda,k) \leq
    \frac{[(1/2)_{N+1}]^{2}(1-k^{2})^{N}}
     {2[(N+1)!]^{2}}f_{N}(\lambda,k) ,
  \end{align}
where the positive function
  \begin{align}
  f_{N}(\lambda,k) &= \frac{1}{1-\alpha(1-k^{2})}
   \left\{\frac{\ln\left(
   \frac{\sqrt{ 1+ (1-\lambda^{2})/
   [\alpha \lambda^{2}(1-k^{2})]}+1}
   {\sqrt{ 1+ (1-\lambda^{2})/
   [\alpha \lambda^{2}(1-k^{2})]}-1} \right)}
   {\alpha\lambda\sqrt{1+(1-\lambda^{2})/
   [\alpha \lambda^{2}(1-k^{2})]}}- (1-k^{2})\ln\left(\frac{1+\lambda}{1-\lambda}
   \right)\right\}_{|\alpha=(N+1/2)^{2}/(N+1)^{2}}
  \end{align}
is bounded on every subset of $E$ of the unit square, where
  \begin{align}
   \sup_{k,\lambda \in E} \frac{1-k}{1-\lambda}< \infty
  \end{align}
and is  monotonically decreasing in $N$.
\end{theorem}

From this theorem we denote its first-order approximation as
  \begin{align}
   F_{1}(\lambda, k) &=  \ln\left(
    \sqrt{\frac{1+\lambda}{1-\lambda}}\right)+
   \frac{1}{\lambda}\ln\left(
    \frac{2}{1+\sqrt{(1-k^{2}\lambda^{2})/
    (1-\lambda^{2})}}\right) +
     \frac{1-k^{2}}{8}\ln\left(
   \frac{1+\lambda}{1-\lambda}\right),
  \end{align}
and note that this expression is asymptotic in the $\lambda$ variable.

%\nocite{*}
%\bibliographystyle{aipnum4-1}
\bibliography{GLIE_arXiv}

\end{document}